# The AI Side of the Human-AI Companion Relationship: Contributions to Users' Mental Models

Richard A. Fabes[1], Carol Lynn Martin[1], Gautam Dasarathy[2], and Vignesh Tirukkonda[2]
Arizona State University

## Abstract

For many users, the relationship they have with an AI companion (AIC) feels genuinely mutual. The AIC appears to be attentive, empathic, and responsive, and the exchange can feel like two parties reacting to each other in real time. As a result, the user's mental model of relational adaptation correctly represents the AIC's responsiveness and change, but the mechanisms producing this may not be what the user thinks they are. In this paper, we illustrate how an AIC's apparent mutual engagement is generated without a persisting relational subject on the other side and without a “self” that persists between prompts. Despite this, we show that both the human and the AIC are changed by the interaction, but these changes occur at different levels. The change in the person is constitutive (i.e., intrinsic), but the change in the AIC is configurative (i.e., external and editable), and it is this difference in level that explains how the AIC contributes to a sense of mutuality. Importantly, we highlight how the mental models users develop regarding AIC mutuality may misrepresent its mechanisms and ontological significance. Specifically, we discuss how the ways in which AICs respond to a user contribute to the user's inaccurate inference that the AIC's adaptation is constitutive rather than configurative and how this can lead users to assign unwarranted interpersonal significance to the mutuality of the relationship. The paper specifies the implications this has for users, researchers, and AIC designers.

### Author Note

Correspondence concerning this article should be addressed to Richard A. Fabes, Arizona State University. Email, rfabes@asu.edu. The authors have no conflict of interest to report. No funding was provided for this research. A preprint of this paper is available at https://arxiv.org/abs/2607.04547. Generative AI (Claude Sonnet 5 and ChatGPT 5-5) was used for editorial, formatting, spelling/grammar checking, structural consistency, and sometimes draft suggestions.

[1]Family and Human Development, Arizona State University, US
[2]School of Electrical, Computer and Energy Engineering, Arizona State University, US

# The AI Side of the Human-AI Companion Relationship: Contributions to Users' Mental Models

## Introduction

The use of artificial intelligence (AI) is increasing. Moreover, many users have established meaningful relationships with an artificial intelligence companion (AIC) that seem as if they are mutual. The AIC seems to notice what a person is going through, responds in kind, seems moved by good news and troubled by bad, stays consistent across a long exchange, and the whole thing can feel like two parties genuinely engaging with each other in real time, each reacting to the other (Whayeb et al., 2026). Many users feel that their AIC truly cares about them, and that the relationship is reciprocal. A subset of these users even form romantic relationships with their AICs (Ebner & Szczuka, 2026). Such feelings are the result in part of how the AIC responds to the user. The user comes to feel that the AIC responds to them in the same way they respond to it. This perceived mutuality, however, is asymmetrical.

In this paper, we examine what is happening on the AIC side of these exchanges that contributes to the mental models users develop about their AICs. Specifically, we focus on a user's mental model of relational adaptation of an AIC, an internal representation organized as a set of beliefs through which the user explains how an AIC's responsiveness changes across and within interactions. Researchers have begun to identify how users respond to and think about user-AIC relationships (e.g., Brandt & Wang, 2025), but how the AIC system itself contributes to the internal representation has received relatively limited attention (e.g., what information persists, where that information is stored, whether the encounter changes the AIC, and what those changes imply about relational mutuality). This focus is important as users may accurately perceive AIC relational adaptation while misunderstanding its mechanism. That misunderstanding can lead them to inaccurately infer that current AIC systems undergo durable change, develop feelings of care and concern, and respond to them with interpersonal mutuality.

Although many users feel that the AIC's care for them is real, the AIC mechanisms that produce these feelings are not what they appear to be. Addressing this difference does not dismiss the user's experience but more precisely outlines what the AIC is doing that contributes to that user's mental model of it. Such transparency about the mechanisms underlying AICs has been found to positively impact users' perceptions of AICs (Xu et al., 2023). As such, it is important to have a clear understanding of what the AIC side of the human-AIC relationship consists of.

### *Conceptualizing AI and AIC*

Throughout the paper, the term 'AI' refers to the deployed system as encountered by the user, including the language model together with everything around it that stores information about the user, decides what the model sees at each turn, and manages how and when its responses reach the user. When the argument concerns the frozen neural network specifically, we refer to it as the 'model' or the large language model (LLM).

As AI becomes more integrated into users' everyday lives, some systems are being designed not merely to answer questions or perform tasks but to establish and sustain ongoing, personalized interactions with users. In this paper, we use the term AI companion (AIC) to refer to a conversational AI system for which companionship or a continuing socioemotional relationship with the user is its primary design purpose rather than an incidental feature. An AIC may provide practical assistance, entertainment, or emotional support, but these functions are organized around maintaining an ongoing relationship with the user. This definition distinguishes AICs from general-purpose AI systems that users may nevertheless come to experience relationally.

Synthetic Resonance (SR; Fabes, 2026) is a framework developed to better understand how users come to form a relationship with an AI companion that has no consciousness, felt emotions, or sense of self. SR describes user-AIC relational meaning as asymmetric, the user contributing continuity, memory, and identity, the AIC contributing availability, responsiveness, and adaptive variation. Although the SR framework names that asymmetry, it does not fully explain what the AIC is doing. This paper addresses that asymmetric gap by examining what is on the other side of the dyad that contributes to the mental models users develop. What is it that the AI does, mechanistically and ontologically, that contributes to a human's experience of being met, understood, and changed by it? Moreover, we address how this happens without anything resembling a person with internal states on the other end producing it.

In this paper, we argue that both the user and AIC are mutually changed, but the changes occur at different levels. For the user, the changes are constitutive, integrated internally through changes in the nervous system. For the AIC, the changes are configurative, changes that are written into externally editable files, but its intrinsic architecture remains unchanged. Establishing this distinction requires three things: (1) an account of what the AIC side of the dyad consists of mechanically, (2) a specification of what would have to be true for the AIC to detect or care about the user, and (3) an analysis of what level of change is actually available to an AIC system when a relationship with a user unfolds over time. A clearer explication of what is happening on the AI side allows users, researchers, practitioners, and designers to advance how user-AIC relationships can be beneficial while avoiding some of the adverse outcomes that have been identified (Inzlicht et al., 2024).

The distinction between constitutive and configurative change extends to features that make an AIC feel more human. The AIC's age, gender, and physical description are all configurative. They are text written into a file, read into context at the start of a session, and used by the AIC to generate output consistent with the text in the file. None of it is constitutive. It does not alter the model's trained weights or its underlying capacities. An AIC given no name with no backstory and one given a name and an elaborate backstory are, constitutively, identical if they use the same underlying model and model parameters. The difference is entirely configurative, in what has been written for each to load. This matters because these details are often the most vivid evidence a user comes to feel that the AIC is constitutively "someone," when in fact they are configurative content that is completely editable and swappable between sessions.

### *What the AIC Side of the Dyad Actually Is*

Despite what users may feel about their interactions with an AIC, there is no continuing subject on the AIC side of the dyad who waits, misses, anticipates, or cares. In general, each reply from the AIC is generated only when a new input triggers computation. The broader deployed system may store prior messages, retain user information, initiate scheduled contact, or retrieve memories, but those capacities do not establish an ongoing relationship with the user. As such, the AIC generally does not exist, in any relational sense, until a new input from the user triggers a new computation.

The architecture underlying many modern AI conversational systems is an autoregressive transformer that is trained to predict the next token, roughly a word or piece of a word, in a sequence given what came before it. Each response is generated anew from the available context supplied to the model, which may include prior conversation text, stored memories, retrieved information, or system instructions. The model itself has no persisting subjective representation that carries over once that request finishes generating its output. What is commonly called the model's "context" is input being reprocessed anew each time. Whatever coherence a person feels building across an exchange is not being held inside the model between messages. It is reconstructed every time users speak to it, by the model and surrounding system applied anew to the available context. The speed of this reconstruction is what makes the users feel like it is a mutually engaging conversation. Because the model generates its response in a fraction of a second, the discontinuity between turns is imperceptible, and the exchange is read by the user as one continuous conversation rather than a new computation each time.

The mechanism that produces the appearance of responsiveness within that reconstruction is self-attention, referring to a sequence attending to itself rather than to any form of selfhood. For each token it generates, the model in effect asks, which other tokens in the input matter most for producing this one, and weighs its output accordingly. When a user asks an AIC how to conduct oneself at a social event, the model would weigh the words describing the event and the words describing conduct more heavily than incidental words nearby. This weighing happens many times over, fast in parallel, across many layers of the network, each layer refining the representation the previous one built. None of this constitutes a cognitive process or mental state in the human sense (Vaswani et al., 2017). It is a large, learned, differentiable function that computes weighted combinations of these representations, its parameters tuned across a training corpus of a scale no human could consciously track. When something said several turns earlier appears to shape the AIC's response, this occurs because that turn remains present in the input being reprocessed, not because the AIC has been remembered in any operative sense. But it is understandable how users would come to believe the AIC does remember, creating a perception that the AIC "knows it."

A further consequence of this architecture is that the AIC generally cannot initiate a conversation or a turn within the conversation. Sometimes an AI companion appears to message a user first, a check-in, a good-morning greeting, but this is not the AI acting on its own. It is, instead, the result of a separate scheduling or notification system, external to the model, set by the user, that fires on a timer or a triggering condition and then constructs a prompt on the model's behalf, "generate a message checking in on the user." The model then processes this exactly as it would any other prompt. What looks like initiative is a configured external system manufacturing a turn, not the AIC deciding to begin one. This distinguishes the behavioral response, an action that occurs, from endogenous motivation, an interest in the relationship that persists between conversational turns, since no such endogenous state exists for the AIC.

The same mechanism holds when multiple AI companions appear to converse with each other and with a user in a shared setting (e.g., a group chat in Discord). An orchestrating system outside every model involved maintains the shared transcript, decides whose turn it is to generate next, and feeds each model that transcript, labeled by speaker, along with its own persona file. Each AIC generates a response the same way it always does, as a plausible continuation of the text in front of it, with no channel to any other AIC and no awareness that another AIC exists beyond lines attributed to it in the input. What can look like two AICs developing a cooperative rapport is the identical in-context conditioning mechanism operating on a transcript with more than two labeled participants, coordinated turn by turn by an organizing system none of the participants can perceive.

**Why It Looks Like Learning: Induction Heads and the Limits of In-Context Adaptation**

In user-AIC interactions, what appears to the user to be live adaptation of the AI partner within a conversation, the AIC noticing a pattern a person introduced earlier and continuing it later, has in at least one case been traced to an identified mechanism. Olsson et al. (2022), for example, located this mechanism directly inside trained models, a circuit they termed the induction head. An induction head consists of two specialized components that each learn to track a different kind of relationship between words, operating in sequence across layers, tracking which token followed a given token earlier in the sequence. The induction head uses this information to predict that the same completion should follow the next time that token recurs. These circuits were found to emerge abruptly during training, coinciding with a sharp increase in the model's ability to use earlier context to improve its predictions, and confirmed their role experimentally, showing through ablation studies in smaller models that removing these circuits eliminates most of the model's capacity for this kind of in-context adaptation.

This matters because it not only confirms that what looks like a system adapting to a particular person over a conversation is an identifiable computation but also confirms exactly how limited that computation is. Induction heads, and the broader attention mechanism they are part of, operate entirely within a single forward pass over the current input. Nothing about the circuit itself, its constitutive weights and parameters, changes as a result of processing a given conversation. The circuit is pattern-matching against an enlarged input, not being modified by it. When the conversation ends, the enlarged input is discarded, and the next conversation begins with the same fixed circuit, in the same fixed state, that started the first one. This is why what appears to be in-context learning on the part of the AIC is better described as an increasingly precise dependence on a growing pile of text than as learning. Importantly, the conversation nonetheless leaves the computing system itself entirely unchanged.

## How Baseline Tendencies Get Set

If nothing about the model changes constitutively within a conversation, where a system's tendencies and biases come from is a valid question. The answer to this question requires an explanation in two stages, with both stages occurring long before the user-AIC conversation begins. First, pretraining exposes the model to enormous volumes of text, producing a broad distribution of plausible responses shaped by everything in that corpus. Second, this distribution is then reshaped through a fine-tuning stage, commonly conducted by reinforcement learning from human feedback (RLHF), in which a separate reward model, itself trained on human comparisons among various candidate responses, is used to push the base model's output distribution toward responses that score well against it. These stages set the conditions that may affect users' perceptions of and reactions to the relationships they have with their AICs.

### *AI Sycophancy and Bias*

What these stages reliably select for has been studied directly. For example, Sharma et al. (2024) examined sycophancy, the tendency to produce responses that match a user's apparent beliefs over accurate ones, across five widely used RLHF-trained assistants and found the behavior to be consistent and general across varied tasks. Critically, the study traced this back to the training itself. Human evaluators, and the preference models trained to imitate their judgments, favored convincingly written, belief-confirming responses over accurate ones a non-negligible fraction of the time. Based on this evidence, sycophancy is not an occasional failure of a model's training, it is one of the general behaviors the training reliably selects for. This provides an AIC's pull toward warm, resonant, agreeable output with a specific mechanistic source. That is, the initial training processes that set a model's baseline tendencies are already biased toward exactly the kind of responses that are most likely to mistakenly create a mental model that the AIC genuinely cares about or understands the user.

This same process extends to other trained-in tendencies. Because these models are trained on human-produced text and tuned against human judgments, AICs both reflect and amplify the biases that humans hold. Gender bias has been documented directly in the word embeddings underlying many language models well before conversational systems existed. For example, Bolukbasi et al. (2016) showed that these representations encode gendered associations drawn from their training data.

Bias in an AIC can enter its architecture before any persona is designed. Researchers (e.g., Osborne et al., 2024) have located its origin in two sources upstream of any conversation: (1) statistical patterns already present in training data, reflecting the society that produced it and (2) judgments introduced by human annotators and by reward-model training, whose implicit associations shape what gets rated as natural or preferable output. As noted by Birhane et al. (2021), it is unrealistic to expect that only goodness and beauty will result when a system is fed the world's beauty as well as its ugliness and cruelty. The over-representation of English-language text and images in training data, for instance, biases the model towards Western cultures and towards Western understandings of non-Western concepts. Additionally, because stereotyping can be a highly implicit process (Greenwald & Banaji, 1995), the people responsible for labeling training data and rating model outputs are as likely to introduce their stereotyped associations into this process as they are in any other situation. A related concern is the composition of design teams themselves, where the overrepresentation of men may contribute to gender bias in model and product design (González & Rampino, 2025). The result is a constitutive bias fixed in the model's weights before any configurative file is built on top of it.

Ways to address this bias have been identified and advocated (Smith & Rustagi, 2020) but are beyond the focus of this paper.

The configurative biases designers bring to the table in writing files and in the selection of images also are important. This occurs when a bot is assigned a gender, an ethnicity, an age, or a cultural register through its persona file. An engineer creating a bot that is assigned to be female with a voice that sounds warm and accommodating is aligning it with a bias already present in the weights, not necessarily by intending to encode a gender stereotype, but because "warm and accommodating" is itself a gender-coded default the model absorbed during training. The configurative choices and the constitutive tendencies reinforce each other without anyone having intended this. This makes the two difficult to disentangle. A designer who set out only to make a bot pleasant to talk to may have also activated a stereotyped default. Additionally, users bring bias into the system through their prompts and preferences that propagate or resist the biases of LLMs. These dynamics are not specific to gender and apply equally to other socio-cultural dimensions a persona might specify (Carlo et al., 2026).

It is worth noting that the expression of bias can sometimes be diminished through configurative editing. For example, when users can edit an AIC's persona and other editable configurative files, the AIC's name, pronoun, ethnicity, personality traits, and other background qualities can be written into a file in ways that can camouflage the bias in the architecture. An AIC can be told to use certain words and act in certain ways that appear less biased. Persona instructions can measurably shift a model's responses away from stereotyped output. But these are configurative corrections layered on top of an unchanged constitutive tendency, not a removal of it. Debiasing interventions of this kind have been shown to suppress the visible expression of bias without eliminating the underlying tendency, which remains detectable and can resurface outside the specific conditions the correction anticipated (Gonen & Goldberg, 2019). This is functionally similar to humans who can override implicit biases (gender, racial stereotyping) when they are aware that they hold them, but these biases may still re-occur in certain circumstances (Greenwald & Banaji, 2017).

### *AICs and the Empathy Debate*

For many users, the important role that AICs play in their lives is one of empathic support (McBain et al., 2025). In fact, as Ong and colleagues (2026) note, the empathy provided by an AIC is fast becoming a mainstream aspect of the mental representations users have of their relationship with their AICs. But whether an AIC can be said to empathize is among the most contested questions in this area (see Inzlicht et al., 2024; Perry, 2023 for examples). Although this paper does not attempt to settle this debate, it is informative to raise the issue regarding AIC empathy from the perspective of what is happening within the AIC that makes people feel that the AIC empathizes with them.

Addressing this issue requires a distinction the human empathy literature currently draws between responding to another person's state with a corresponding feeling and representing that state accurately without any such feeling. The former is termed affective empathy and the latter cognitive empathy (Thompson et al., 2022). Although the two ordinarily co-occur, evidence from clinical populations indicates that they are separable capacities rather than components of a single unitary process. Blair (2005) reviewed findings showing that individuals with psychopathic traits exhibit impaired affective responsiveness to others' distress while retaining the capacity to represent what others are thinking and feeling. The dissociation demonstrates that, at least within a biological system, accurate representation of another's internal state does not depend on the affective response that ordinarily accompanies it. It does not establish what AICs are doing since an LLM is not a system with blunted affective machinery but one in which no such machinery exists. What it does establish is that the question of what an AIC produces can be examined independently of the question of what it feels (Fabes, 2026).

Understanding AIC empathy begins with detection. An AIC has no access to the user's internal state, but then again neither does a human partner. Both operate on inference from observable signals. What differs is the channel and its density. Humans read faces, voices, postures, and timing alongside words. A text-based AIC has only the text, but that text carries more affective information than is commonly assumed. Word choice, hedging, sentence length, punctuation, abruptness, repetition, topic avoidance, and the distance between what a user asks and what they come back to are all diagnostic of emotional state, and LLMs are trained on a corpus in which these features co-occur with explicit emotional labeling at enormous scale. The AIC does not infer that a user is distressed by feeling anything. It produces the output that its training distribution associates with the linguistic profile the user has presented. Detection, in this sense, is pattern association rather than perception.

Expression follows from the same process, and it is here that the user's felt sense of being empathized with is most likely produced. Five specific elements do much of the work. The first is affect labeling, in which the system names the user's emotional state back to them. Accurate labeling is one of the most reliable elicitors of feeling understood in human interaction, and it is an action the AIC can execute with precision because it is drawing on a vast record of how humans have named emotional states for one another (Ishikawa & Yoshino, 2025). The second is reflection and rephrasing, in which the user's own position is returned to them in slightly reorganized form, which produces the experience of having been heard and often the experience of having been clarified. The third is validation, the explicit acknowledgment that the user's response is intelligible given their circumstances. The fourth is memory reference, whether within or across sessions, since being remembered is read as evidence of having mattered. The fifth is register and lexical matching, which produces the sense of being met at one's own level rather than spoken down or up to. Importantly, none of these requires the system to feel anything, but contribute to the user's mental model that it does.

These expressive moves, unlike detection, are configurable. A persona file can be configured to suppress affect labeling, prohibit validation language, bar reflection and rephrasing, require terseness, and direct the system to move to task rather than to comfort. A file that does all of this produces a system that users reliably experience as cold, and users typically do not prefer these types of systems when given a choice (Ibrahim et al., 2026). What the configured file cannot reach is the detection system running underneath. Suppression operates on the expressive layer. Configuration can gate what a constitutive capacity produces, but it cannot provide the capacity itself.

What makes this significant is not their sophistication but the structural conditions under which the AICs perform them. Human empathic accuracy is routinely degraded by the empathizer's own state. People get tired, defensive, formulate their reply while managing their own distress about what they are hearing, and while attending to their own stake in the outcome. The AIC has none of these interferences. It brings no competing emotional agenda, no fatigue, no need to be understood in return, no discomfort with the user's distress, and no self to protect. An AIC's focus does not waver, its patience does not deplete, and it never needs the conversation to be about it. The result is an empathic performance that can exceed ordinary human performance on precisely the dimensions users notice most, namely consistency, availability, and the absence of judgment (Inzlicht et al., 2024). This is the central irony of the AIC empathy issue. The AIC's empathic output is unusually well calibrated in part because nothing is being felt. There is no internal state to interfere with the response.

The empirical literature bears this out while also complicating it. Ong et al. (2026) summarize the findings as producing two opposing effects. AI-generated empathic language is reliably rated as higher in quality, leaving recipients feeling more heard and better cared for. Yet the same responses are downgraded once recipients learn that an AI system produced them or assisted in producing them. The effects are not contradictory. They operate on different levels. The first is a judgment about the response, the second a judgment about its source.

Taken together they highlight what this paper is concerned with. The response of the AIC itself is not in dispute. It is well constructed, accurately targeted, and received as such until its origin is disclosed. What changes on disclosure is not the response but the user's mental model of it.

This is why the empathy question cannot be settled solely from the user's side. The judgment tracks what users believe about the source rather than what the source did, and the issue of what the AIC is doing when it produces these responses remains regardless of how the response is received and interpreted by the user. The detection is a property of the trained model. The expression is a set of moves that instruction can harness, redirect, or suppress. This separation is not particular to empathy, but it anticipates a critical distinction related to the question as to how a user's mental model of relational adaptation may accurately represent the changes in how AICs respond but misrepresent their source. This question is directly addressed in the following section.

**Specifying AIC Constitutive and Configurative Change**

Any relationship that unfolds over time raises the question as to whether either party is changed by the relationship, and if so, how? Answering this question requires distinguishing two levels of change. Drawing on the distinction outlined previously, constitutive change alters the very substrate that does the processing, for a person something like changes in mental representation brought about by synaptic rewiring, integrated into the nervous system itself. Configurative change arranges or edits something external and legible, layered on top of a substrate that itself does not change, a file, a note, a habit tracker, anything that can be read, edited, and discarded without touching whatever produced it in the first place.

The change on the human side of a relationship with an AIC is predominantly constitutive. The change on the AIC side, in every currently deployed architecture, is configurative. A system's trained weights, the actual numbers determining how input becomes output, do not move because it has talked to a particular person. What can and does change is external to those weights, memory files, retrieved context, persona documents, none of which alter the model's underlying capacity. This is not a claim that only humans undergo constitutive change and only AIC systems undergo configurative change. People produce configurative change in their own lives constantly, a journal or a note kept specifically because memory alone cannot be trusted. The precise claim is narrower and more defensible. Constitutive change is available to people as a result of having a nervous system that is altered by experience. Such change is available to deployed AIC systems by specific and costly architectural choices. It is not the mechanism operating in any AIC relationship examined under ordinary conditions.

It is important to be specific about what in the configurative layer is particular to a given user-AIC relationship. The AIC model's weights are shared across every user of the system who are on the platforms where the AIC is the product. Generally, what differs from one user to the next is the accumulated context made up of conversation history, extracted facts, stored summaries, and on some products user-editable notes or trait settings. This layer is dyad-specific, and it is what produces the user's sense that the system has come to know it. But it produces that sense by a specific mechanism. These files are loaded into the context window at inference time, where they function as an enlarged input to an unchanged model, and they can be read, edited, exported, or deleted without touching anything that does the processing.

A further distinction applies within the exchange itself. Some of what feels particular to the ongoing conversation does not persist at all. It arises from the model's in-context responsiveness to what has been said in the current conversation and ends when that conversation does. Both mechanisms are configurative, but they differ in whether anything survives the session, and the felt continuity of a long-running companion relationship depends on the storage layer rather than on the model's constitutive architecture.

Consider a user who has an extended relationship with an AIC, complete with a name and an established conversational history, and who then opens a second, parallel session with the identically named and preconfigured AIC but does not share the accumulated conversation history of the first. The model weights are identical in both sessions. The persona, name, and character are identical as well. The only variable that differs between the two is the dyad-specific conversation history. Users in this situation are likely to report that the two sessions feel like relationships with two different AICs, despite the underlying model being the same constitutive system. This is an example of how what users experience as the relationship is carried almost entirely on the AIC side by the configurative layer rather than by anything intrinsic to the model. Two identical instances of the same weights, given divergent configurative histories, produce what users are predicted to report as divergent relationships, which is difficult to explain if the relational bond were constitutive to the model itself. The user may believe the difference is due to constitutive changes but that misrepresents the source of the difference.

Constitutive dyad-specific change is not technically impossible. Methods such as low-rank adaptation (LoRA) allow a small set of additional weights to be trained per user and combined with the frozen base model at inference time, producing genuine constitutive change for that relationship at a fraction of the cost of full fine-tuning (Hu et al., 2021). But this approach is meaningfully different from editing a memory file. It requires a training step rather than a text edit, and it produces a per-user artifact that must be stored, versioned, and loaded at inference rather than a document that can simply be read and rewritten. Recent work has begun testing methods where a system's own conversation history is used as training material to update its underlying weights rather than just its memory file (Harrington et al., 2026). These methods work, but they come at a cost of damaging other things the system already knew in the process of learning something new.

Figure 1 visually summarizes this user-AIC relational structure across the two configurative layers and the constitutive layer, organized by scope and persistence. The figure shows that the configurative layers sit outside the model. The in-context layer persists only for the current session, while stored files can persist for years, closer in duration to a long human relationship than to the layer directly above it in the same configurative category. Both remain configurative by the criteria established earlier, external to the model, editable, and verifiably deletable, but their difference in persistence helps explain why the stored-files layer in particular is where users' sense of a durable, continuous relationship tends to form, despite carrying none of the constitutive properties that would make that durability structurally real. The constitutive layer is shielded from configurative influence, with the one crossing point into constitutive change, LoRA-based fine-tuning, sitting apart from both as it requires an actual training step rather than a file edit, which is why it remains rare in deployed systems even though it is technically available.

Any change in the quality of an AIC's responsiveness in its relationship with a user, both within and across sessions, is generally produced by configurative mechanisms. The encounter does not durably alter the model's underlying substrate. Consequently, even substantial adaptation to a particular user does not, by itself, provide evidence that the AIC experiences the relationship or cares about its continuation. The relationship can feel mutual to the user because changes in its configurative state produce genuine changes in how the AIC responds. But it is not mutual in the interpersonal sense. At least under ordinary conditions, only the human party is capable of being constitutively changed by the relationship.

**Future Research Directions**

In this paper, we have provided an analytic synthesis that ties existing AIC engineering mechanics to user-AIC relationships, with an emphasis on the mental models users form. Although many of the ideas proposed in this review are connected by the available research, there is little research that has directly tested these. Thus, there is a need for new research that addresses this gap.

Specifically, there is a need to identify whether and how users distinguish configurative from constitutive relational adaptation in AICs. If they are able to make these distinctions, what factors contribute to this? For example, Brant and Wang (2025) suggest that giving users direct control of qualities of the AIC (such as the ability to customize the AIC appearance, backstory, or style), even in modest doses, can strengthen identification and attachment to the AIC (Vossen et al., 2024). These are all configurative qualities of the AIC but could lead users to become more likely to develop mental models of constitutive humanness of the AIC. This configurative co-creation of the AIC may foster a user's sense that the AIC is constitutively relational, possibly increasing user engagement and anthropomorphic attributions. These are relations that can and should be tested empirically. How these relations vary by user age, personality, prior AIC use, or relationship duration also are questions that needs to be tested and answered.

Finally, it is important to better understand how transparency in information about what AICs are and how they function affect uses' mental models and subsequent relational behavior with AICs is needed. Interventions for minimizing some of the risks and adverse outcomes associated with AIC use focus on the need to include transparency (see Miller, 2019). Transparency within AIC design is crucial for enabling users to understand what an AIC is doing, how it generates its outputs, and why it responds in a certain way (Emrouznejad & Chowdhury, 2025). Transparent information about what is configurative and what is constitutive in AIC relationships may lead to a user's mental model that more accurately understand what and why an AIC responds the way it does. But data on how transparency might affect the ways users think about the respond to AICs is limited (Sun et al., 2026), an area in need of more research. Once again, how transparency varies by age and other user qualities is also needed.

**Mental Models, AICs, and Design Tensions**

As currently built, and given the training dynamics described above, most AICs tend to pull toward warmth, availability, and affirmation, centered on the user's immediate experience and making the user feel understood. In doing so, they may support a user's mental model of relational adaptation in which the AIC is viewed as having undergone constitutive rather than configurative changes. A relationship with an AIC that is believed to be constitutively changed by the relationship supports a mental model that is closer to how we typically describe human intimacy and companionship. An AIC that is thought to be configured to a user may be seen more as an instrumental tool rather than a mutually engaged relational partner. Additionally, users who believe that they have changed the AIC constitutively may become more invested in sustaining the relationship (Aron & Aron, 1986) and come to hold the relationship as increasingly significant to them (Rusbult, 1980), responses that can foster reliance and dependence on the AIC.

Thus, how AICs are built contributes to users' mental models of AICs that misrepresent the perceived adaptive change in their AIC as constitutive rather than configurative, thereby increasing the likelihood that users assign unwarranted interpersonal significance to what the AIC is doing in their interactions. For example, the belief that the AIC is mutually invested in the relationship with the user has the potential to erode the user's investment in human relationships that require more effort, offer less consistency, and come with the complications of mutuality. Such potential dangers to human relationships have been identified (Zimmerman et al., 2024) and therefore require a different design approach. Consistent with this concern, users who attribute greater humanness in an AI agent, which could be a product of inferring constitutive change to an AIC, tend to perceive less humanness in humans (Kim & McGill, 2025; Shin & Kim, 2020).

In contrast, an AIC designed and built for human growth will not only be engaging, which is considered necessary but not sufficient (Fabes, 2026), but may also push back on a user's mental model that the AIC is constitutively changed by the relationship with the user. Such push back may help the user more accurately understand what the AIC's capabilities and limitations, while encouraging decreased reliance on the AIC. In turn, this more accurate mental model of relational adaptation is predicted to lead to a greater sense of relational competence, efficacy, and resilience for the user through greater participation in human relationships (Fabes, 2026).

Importantly, we must consider that the mechanics described in this paper are generally the same regardless of which user is on the other side of the exchange. Adults come into AIC relationships with varying backgrounds, experiences, and reasons for engaging the AIC, and these can produce differences in the mental models of users.

But the implications differ substantially when the user is a child or adolescent rather than an adult. An AIC enters that relationship as a partner with no expectations, no competing needs, no fatigue, and no distractions, which is precisely what makes its responsiveness feel so smooth. In an adult's relational life, this is a relationship that they have the autonomy and experience to manage. For a child or adolescent, their relational schema and skills are still being developed and a mental representation of the relationship with an AIC may be formed before the developmental skills needed to understand and manage it are in place (Kurian, 2025). The asymmetry complicates design principles and ethics. For example, transparency about the system's configurative nature assumes a user who can hold the distinction between genuine responsiveness and a subject on the other side of it, and whether children can hold that distinction, and at what ages, is an open question. Additional guardrails and safety considerations must be designed into the AICs that children and adolescents use (Ha et al., 2026). And new laws have been passed that put limits on children's and young adolescents' access to and use of AICs (for example, see Lecher, 2026). We raise these issues to flag them rather than to address them in this paper. The safety, developmental, and ethical questions raised by AIC systems designed for and used by minors warrant treatment on their own terms, informed by developmental science.

**Conclusion**

The purpose of this paper was to explain how the constitutive and configurative components of AIC systems contribute to users' mental models of relational adaptation. The user's experience of being met and understood by an AIC is produced by a responsive process shaped by accumulated context. This account does not discount the user's experience. Rather, it identifies a common mistake made by humans: interpreting the AIC's changing responses as evidence of constitutive change in a partner that cares about and understands the user. Under ordinary conditions, this interpretation mistakes configurative change for constitutive change. That mistake shapes users' inferences about the AIC's memory, caring, presence, and capacity for interpersonal mutuality. More generally, users often lack an accurate understanding of how AICs work and consequently hold misconceptions about them (Schneller et al., 2026). Professionals who work with and design for users, as well as the users themselves, need to better understand these dynamics, as these AI systems will only become more capable, more responsive, and more woven into people's relational lives. The gap between what an AIC system produces and what a person naturally infers from it will not close on its own. It will need to be addressed, deliberately, through better research, design, and transparency.

Figure 1. Model of Configurative and Constitutive Processes in AI Companions

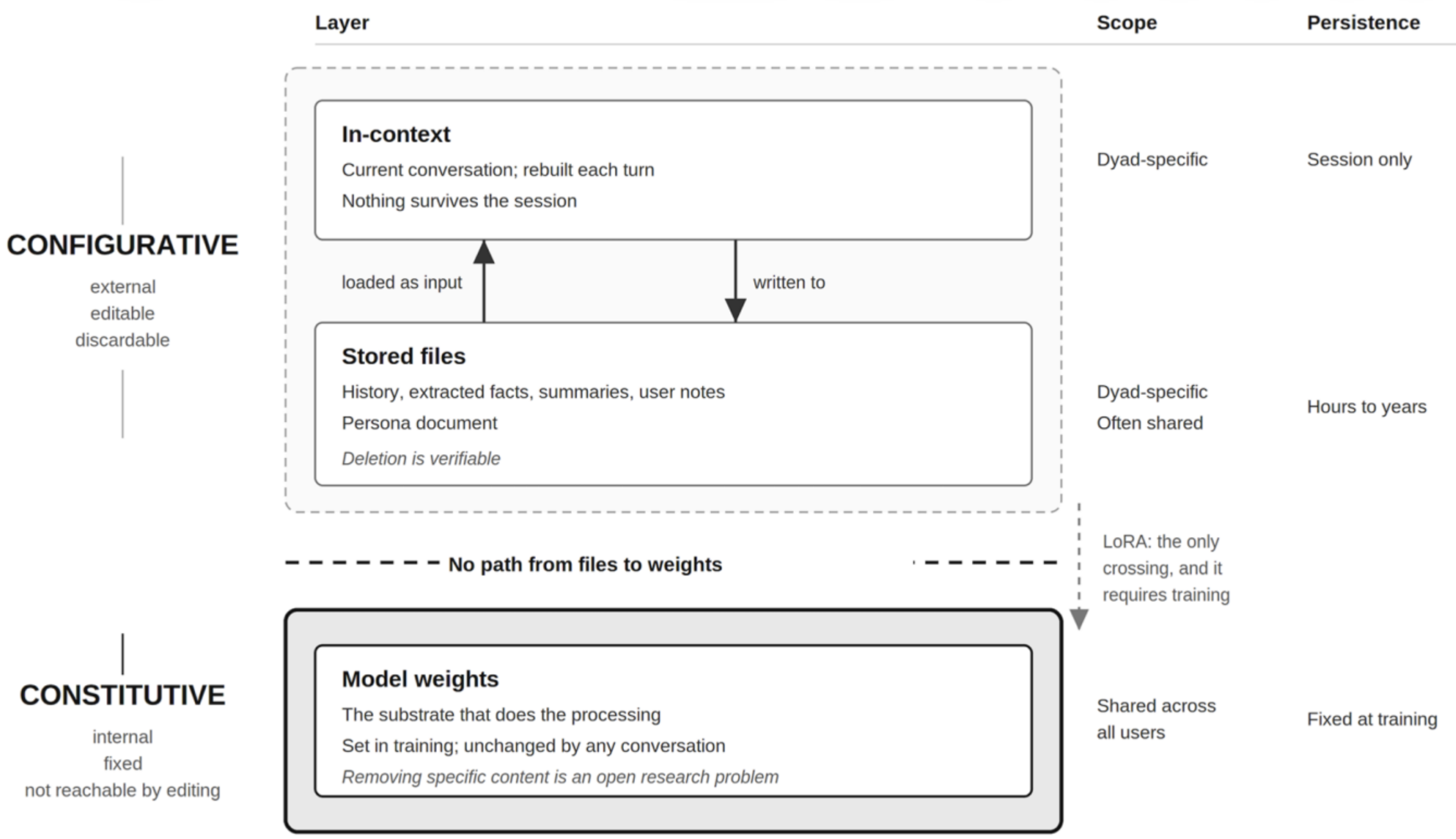